\documentclass{article}

\usepackage[final]{neurips_2025_ml4ps}

\usepackage[utf8]{inputenc} 
\usepackage[T1]{fontenc}    
\usepackage{hyperref}       
\usepackage{url}            
\usepackage{booktabs}       
\usepackage{amsfonts}       
\usepackage{nicefrac}       
\usepackage{microtype}      
\usepackage{xcolor}         

\usepackage{graphicx}

\usepackage{siunitx}
\usepackage{bigstrut}
\usepackage{multirow}

\usepackage{natbib}
\title{DNN-based Signal Processing for Liquid Argon Time Projection Chambers}

\author{%
  Avinay Bhat \\
  Department of Physics, University of Chicago \\
  Chicago, IL 60637 \\
  \texttt{abhat@uchicago.edu} \\
  \And
  Mun Jung Jung \\
  Department of Physics, University of Chicago\\
  Chicago, IL 60637 \\
  \texttt{munjung@uchicago.edu} \\
  \And
  Gray Putnam \\
  Fermi National Accelerator Laboratory \\
  Batavia, IL 60510 \\
  \texttt{gputnam@fnal.gov} \\
  \And
  Haiwang Yu \\
  Brookhaven National Laboratory \\
  Upton, NY 11973 \\
  \texttt{hyu@bnl.gov} \\
}

\begin{document}

\maketitle

\begin{abstract}
  We investigate a deep learning-based signal processing for liquid argon time projection chambers (LArTPCs), a leading detector technology in neutrino physics. Identifying regions of interest (ROIs) in LArTPCs is challenging due to signal cancellation from bipolar responses and various detector effects observed in real data. We approach ROI identification as an image segmentation task, and employ a U-ResNet architecture. The network is trained on samples that incorporate detector geometry information and include a range of detector variations. Our approach significantly outperforms traditional methods while maintaining robustness across diverse detector conditions. This method has been adopted for signal processing in the Short-Baseline Neutrino program and provides a valuable foundation for future experiments such as the Deep Underground Neutrino Experiment.
  \end{abstract}

\section{Introduction}

Recent efforts in neutrino physics aim for increasingly precise measurements of neutrino properties, and liquid argon time projection chambers (LArTPCs) have emerged as a scalable, high-resolution detector technology that enables such precision~\citep{SBNProgram, DUNE}. LArTPCs detect the ionization electrons from the trajectories of charged particles traversing liquid argon, which are drifted by electric field to planes of charge sensing wires (typically three). The inner induction planes measure the induction current from the drifting electrons, while the outermost collection plane collects the charge. Combined with timing information, the 2D wire plane readouts, each representing projections at different angles, enable full 3D reconstruction of particle trajectories. The image-like structure of these data naturally motivates the use of machine learning (ML) techniques~\citep{SPINEML, MicroBooNEML, MLReview}. 

Efficient and robust signal processing of the wire plane readouts is essential for managing the high data throughput in LArTPCs. 
Signal processing on induction planes is especially challenging due to the signal cancellation caused by the bipolar signal response, especially for signals that are extended in time. Furthermore, various detector effects in real data, such as argon impurities, gain variations, and malfunctioning wires, need to be addressed for unbiased signal reconstruction~\citep{ICARUSCalibration}. 

\begin{figure}
  \centering
  \includegraphics[width=\textwidth]{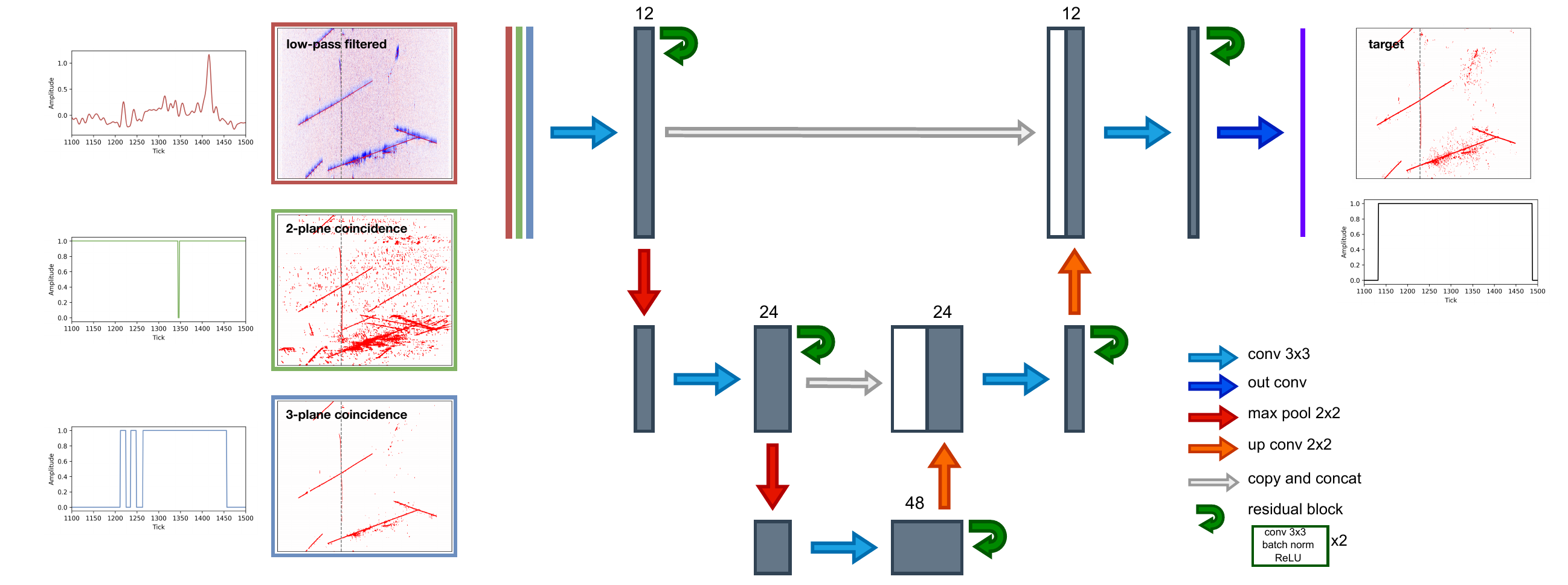}
  \caption{Training scheme using a 3-layer U-ResNet architecture. Input channel images are constructed by applying a low-pass frequency filter and identifying cross-plane coincidences. Example input and target images and waveforms are shown for an elongated track, a case which the signal reconstruction is challenging.}
  \label{fig:network}
\end{figure}

A key step in LArTPC signal processing is the identification of charge signals on individual waveforms as regions of interests (ROIs). Traditionally, this is performed with a thresholding algorithm~\citep{WireCellSP}. However, such an approach does not fully exploit the geometric and topological information available in LArTPC data. In this work, we explore using a deep neural network (DNN) method for ROI identification that leverages both the inherent 2D nature of the data and the correlations between wire planes. We apply this method to the Short-Baseline Neutrino (SBN) program LArTPCs: SBND and ICARUS~\citep{SBNProposal, ICARUSInagural}.

\section{Deep Neural Network for ROI Identification}
\label{sec:DNNROI}
We approach ROI identification as a segmentation task using three-channel 2D image inputs, where each image has wire number and time tick as the axes, extending the framework introduced by~\citet{DNNROI}. The first input channel is the deconvolved signal on the target plane with a low-pass filter applied to suppress noise, where the pixel value is the induced charge. The second and third input channels are binary masks indicating time-window coincidences between the other two planes and among all three planes, respectively. These plane-coincidence masks help the network disambiguate true physics signals from noise and artifacts that are unlikely to be correlated across planes. This framework is depicted in Fig.~\ref{fig:network}.

Training samples were generated from a simulation of particle interactions, propagated through detector models in SBND and ICARUS. We generated samples representative of neutrino interaction events expected at the detectors~\cite{GenieUserManual}, overlaid with a cosmic ray simulation~\citep{Corsika}. In addition, we enriched our sample set with events that are challenging from a signal processing perspective: shallow-angle muon tracks and electromagnetic showers, which are prone to signal cancellation due to the bipolar signal on induction planes.

We tested three U-Net variants: U-Net~\cite{unet}, U-ResNet~\cite{uresnet1, uresnet2, uresnet3}, and Nested U-Net~\cite{nestedunet}. Each was optimized with hyperparameter scans over learning rate, optimizer, and regularization. Due to the sparsity of LArTPC images, input images have a strong class imbalance between ROI and non-ROI pixels. To mitigate this, the loss function was computed with pixel-wise weighting, assigning a weight of 9 to ROI pixels and a weight of 1 to non-ROI pixels. 

A threshold of 0.5 was applied to the predicted scores to classify each (wire, time tick) pixel as either ROI or non-ROI, and the performance was evaluated using the Dice-S{\o}rensen coefficient to pick the best model for each wire plane of each detector. For both SBND and ICARUS, the best model was a U-ResNet, an architecture that replaces U-Net's convolutional blocks with the residual blocks from ResNet~\cite{resnet}. The best SBND model was trained with the SGD optimizer with learning rate 0.01, momentum 0.9, and decay weight 0.0005. The best ICARUS model was trained with the ADAM optimizer with learning rate 0.001, $\beta_1$ 0.9, $\beta_2$ 0.999. Models were trained on NVIDIA A100 40GB GPUs with batch size 8. Code is available at \url{https://github.com/wjdanswjddl/Pytorch-UNet.git}.

Inference was run on CPUs due to the availability of computing resources. To satisfy computing constraints, the input image sizes were reduced by downsampling and chunking. Along the tick axis, we applied downsampling by averaging over fixed intervals of ticks. Along the wire axis, images were split into smaller chunks to allow for inference to run on smaller images, and later stitched together to reconstruct the full output. ICARUS applied tick sampling by a factor of 4, used 2 chunks for the first induction plane (image size of 1056$\times$1024 pixels), and 4 chunks for the second induction plane (image size of 1400$\times$1024 pixels), SBND used tick downsampling by a factor of 4 and used 2 chunks (image size of 992$\times$857 pixels) for both induction planes. Inference took 30s wall time and 3GB memory for SBND and 700s and 8GB for ICARUS per event.

\section{Detector Variation Samples}\label{sec:detvar_samples}

\begin{table}
  \caption{Parameter distributions used for detector variations. \textit{Unif} denotes a uniform distribution, and $N$ denotes a normal distribution.}
  \label{sample-table}
  \centering
  \begin{tabular}{llll}
    \toprule
    Detector   &  Parameter     & Nominal Value     & Variation Distribution \bigstrut\\
    \midrule
    \multirow{6}[2]{*}{ICARUS} & Coherent Noise Scale & 1 & $N(1,0.05^2)$ \bigstrut[t]\\
    & Incoherent Noise Scale & 1 & $N(1,0.05^2)$\\
    & Electron Lifetime & \SI{5}{\milli\second} & \textit{Unif}(2,10) \SI{}{\milli\second}\\
    & Relative Gain (per-plane) & 1 & $N(1,0.1^2)$\\
    & Shaping Time (per-plane) & \SI{1.3}{\micro\second} & $N(1.3,0.05^2$) \SI{}{\micro\second}\\
    & Middle Ind.~Signal Shape & Bin 7 & Bins \textit{Unif}(1-15) \bigstrut[b]\\
    \midrule
    \multirow{2}[2]{*}{SBND} & Smear Waveforms & 1 & $N(1,2^2)$ \bigstrut[t]\\
    & Pixel weight & 1 & $N(1,0.05^2)$ \bigstrut[b]\\
    \midrule
    Common & Masked Region Width & 0 & $N(10,5^2)$ wires \\
    \bottomrule
  \end{tabular}
  \label{tab:detvar}
\end{table}

\begin{figure}
  \centering
  \includegraphics[width=\textwidth]{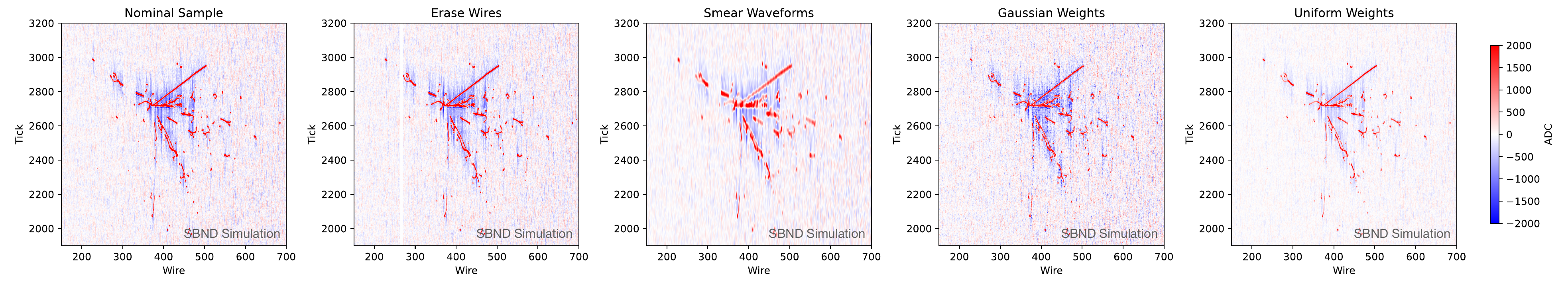}
  \caption{Image augmentations used to mimic detector variations, exaggerated for clarity. From left to right: nominal image, image with masked wires, image with tick-direction smearing, image with pixel-wise random scaling, and image with event-wise random scaling.}
  \label{fig:sample_aug}
\end{figure}

Signal processing is the first step in the data analysis chain, and its stability against detector effects is crucial. The test of resilience against variations in the detector performance represent an example of ``scientific robustness'' in the network training~\citep{SciRobust3, SciRobust2, SciRobustNhan}. We explore two strategies for incorporating realistic detector effects. For ICARUS, detector variations accounting for perturbations in the noise level, argon purity, and ionization signal gain and shaping were simulated through the full detector simulation~\citep{ICARUSCalibration}. For SBND, variations were applied as image augmentations. Smearing effects (e.g., ionization diffusion) were modeled by convolving images with a gaussian kernel, while scaling effects (e.g., gain variations) were modeled with pixel-wise and event-wise scaling of pixel values. Examples of such augmentation are shown in Figure \ref{fig:sample_aug}. To emulate malfunctioning wires, bands of wires were randomly masked in the training input images, allowing the model to learn the general behavior rather than memorize the specific bad regions. These variations are summarized in Table \ref{tab:detvar}

To evaluate the robustness of DNN ROI against detector variations, we generated test samples for ICARUS by simulating key variations observed in the detector: high noise, low electron lifetime, and increased wire intransparency. These variations are extreme with respect to the operation of the ICARUS detector, and should therefore be considered on the edge of the possible detector performance. For example, the low electron lifetime applies a value of \SI{2.5}{\milli\second}. This value attenuates the charge signal by up to 33\%, and is lower than any value included in ICARUS physics data~\cite{ICARUSCalibration}. As part of this publication, we release a sample of ICARUS events with the detector variations to provide a dataset relevant for scientifically robust image identification for particle physics applications, detailed further in the Appendix.

\section{Results}\label{sec:results}

\begin{figure}
  \centering
  \includegraphics[width=0.485\textwidth]{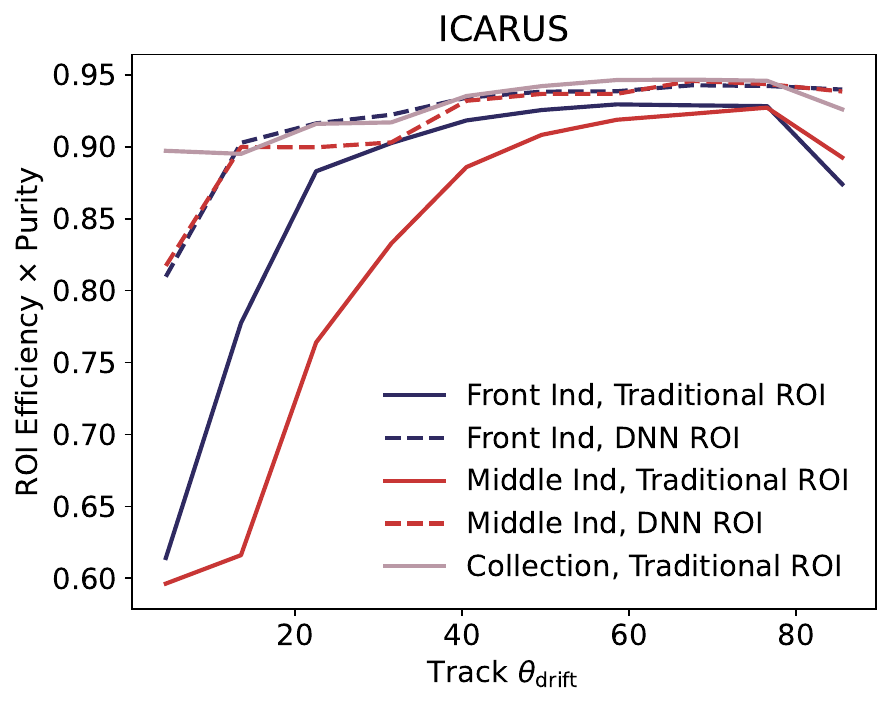}
  \hspace{5pt}
  \includegraphics[width=0.46\textwidth]{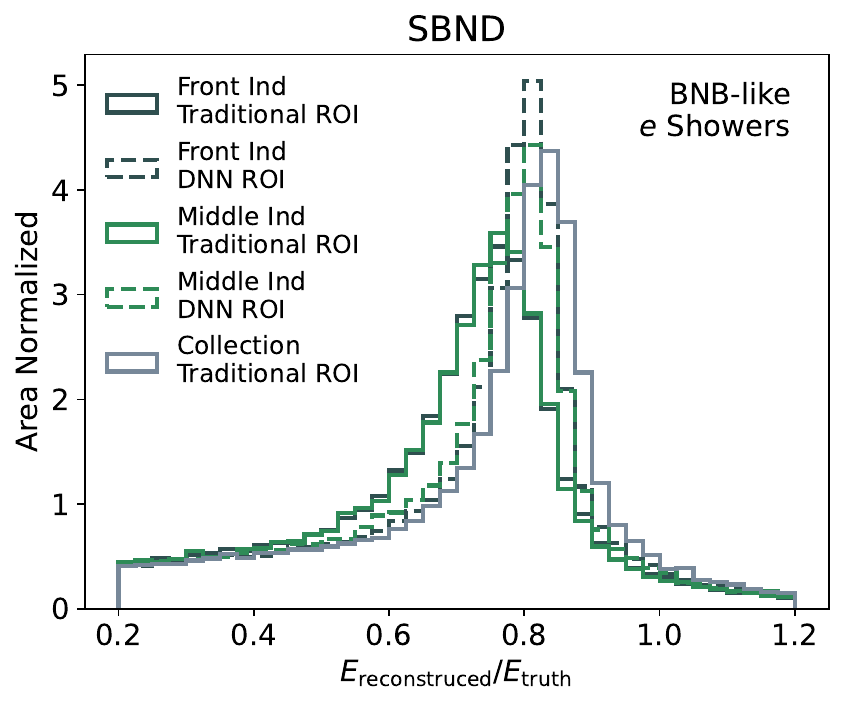}
  \caption{ROI efficiency and purity for track-like charge depositions in ICARUS, shown as a function of the track angle relative to the drift field ($\hat{x}$). Results are based on 10,000 test samples, with a charge threshold of $10^4$\si{e}$^-$. Right: Reconstruction performance for BNB-like electromagnetic showers in SBND, evaluated on 20,000 test samples.}
  \label{fig:results-showers}
\end{figure}

As shown in Figure \ref{fig:results-showers}, DNN ROI significantly outperforms the traditional thresholding algorithm on both SBND and ICARUS. ROI identification for track-like deposits, such as muons and protons, improves across all angles, with the largest gains for elongated tracks. Enhanced energy reconstruction of neutrino-induced electron showers further confirms the benefit of improved ROI finding on reconstruction of physics variables. This improvement arises from two factors: DNN ROI more efficiently recovers attenuated signals from portions of showers that are oriented at elongated angles and better identifies small, isolated charge deposits, leading to more complete shower energy reconstruction. The reduced dependence of performance on specific particle properties like track direction or shower energy improves the consistency of ROI finding across a wider range of particles. 

\begin{figure}[t]
    \centering
    \includegraphics[width=0.46\textwidth]{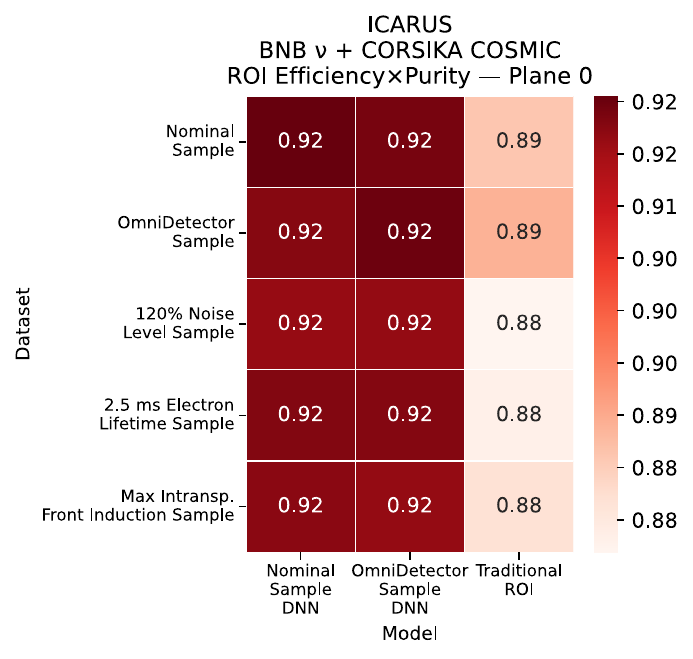}
    \includegraphics[width=0.46\textwidth]{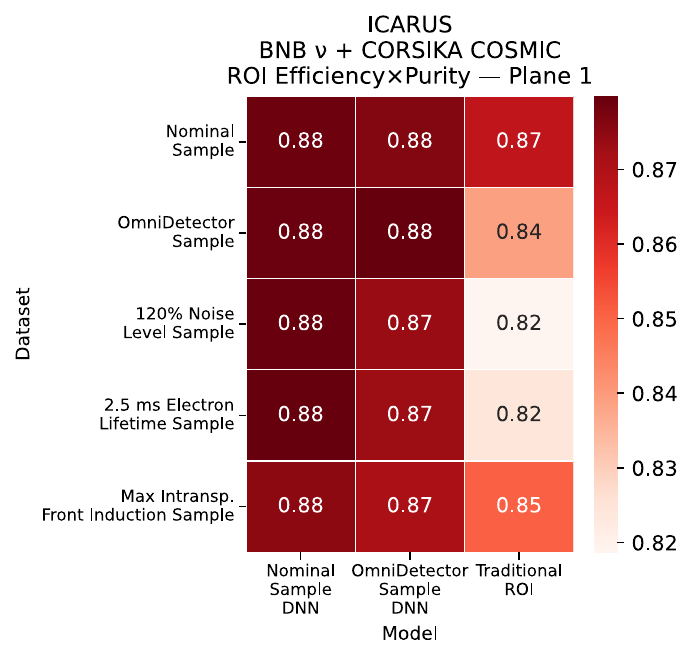}
    \caption{ROI Efficiency$\times$Purity for plane 0, Right: for plane 1 in ICARUS, from simulated neutrino interactions in BNB with simulated cosmic rays. Results are shown under different detector variations, comparing traditional ROI and DNN ROI models.}
    \label{fig:icarus-var-bnb}
\end{figure}

\begin{figure}[t]
  \centering
  \includegraphics[width=\textwidth]{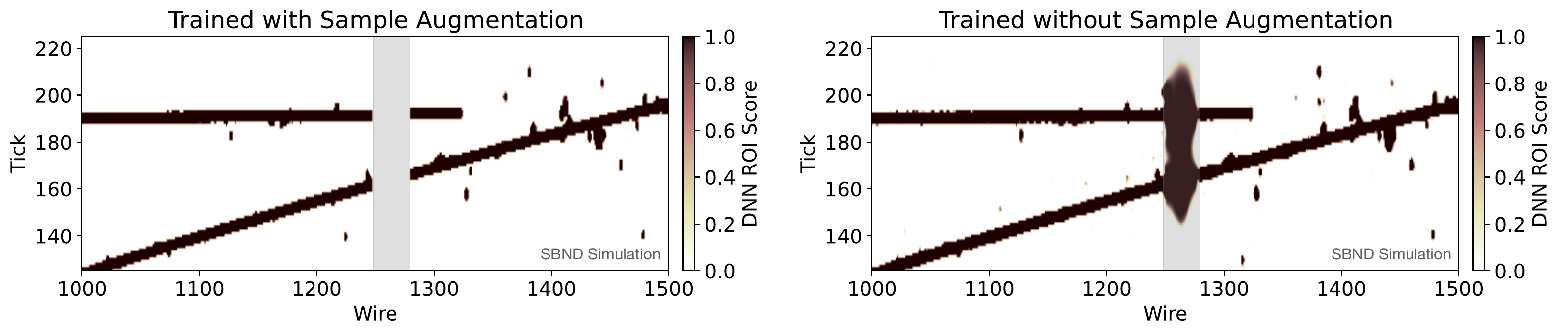}
  \caption{Left: DNN ROI score prediction from a network trained with augmentation, Right: without augmentation, applied to an SBND event. The gray band indicates a detector dead region.}
  \label{fig:results-deadchannels}
\end{figure}

In addition to improved ROI identification performance, DNN ROI shows significantly greater robustness to detector effects. Figure ~\ref{fig:icarus-var-bnb} shows the comparison of the ROI identification performance of two DNN models, trained with and without detector variation samples, against the performance of the traditional method in ICARUS, using the ROI efficiency$\times$purity figure-of-merit on detector variation test samples described in Sec. \ref{sec:detvar_samples}.

The traditional method is particularly sensitive to variations on the middle induction plane, where bipolar cancellation is strongest. For example, increasing the noise by 20\% on this plane (a deviation well outside the expected range: four standard deviations from the average simulated noise scale) reduces the efficiency$\times$purity metric by 7.5\% relative to the nominal simulation. In contrast, DNN ROI maintains stable across all test samples up to statistical noise in the results, even without training on variation data. This robustness is likely due to the network's ability to identify pattern matching features in its input images, informed by the cross-plane coincidence inputs. These features may be inherently more robust against detector effects than the signal-to-noise thresholding applied by the traditional algorithm. Furthermore, the diversity in the topology and magnitude of charge depositions from different particle types, energies, and orientations present in each training sample may already represent a broader set of signals than are induced by variations in the detector simulation.

While inclusion of augmented training samples does not meaningfully improve overall figure-of-merits, such augmentation is essential for handling certain data defects. For example, Fig. \ref{fig:results-deadchannels} shows that a network trained on augmented samples correctly ignores dead wires, whereas a nominally trained network predicts unphysical ROIs in these regions, although the confusion doesn't affect predictions elsewhere. As demonstrated, DNN ROI may still behave unpredictably under substantially different detector configurations and may require new types of training samples in such cases. 

Future work could further improve sensitivity to other rare or challenging cases. For instance, performance on lower-energy depositions could be improved by optimizing the charge thresholds used for cross-plane coincidence frames and ROI labeling during training, which were conservatively set here to target track- and shower-like neutrino signals.

\section{Conclusion}
We have explored a DNN-based method for identifying ROIs in LArTPC data and demonstrated its effectiveness on the ICARUS and SBND detectors. This approach leverages the image-like nature of LArTPC data and incorporates cross-plane information for improved signal discrimination. From tests using training samples with realistic detector variations, we find that the network is inherently robust to such effects. DNN ROI outperforms traditional methods in both performance and stability even without explicit variation training, likely due to geometric constraints introduced through the cross-plane information. This method is now adopted for use in the SBN program and enhances its capability for precise neutrino reconstruction. 

The application of DNN ROI in SBN situates in the broader context of ML applications in particle physics (see, e.g.~Ref.~\cite{NOvAML, CMSML, AntimatterML, MicroBooNEML, JetML, NoVAML2, SPINEML, MLReview}). In particular, the upcoming Deep Underground Neutrino Experiment, which will also use LArTPCs to study the properties of long-baseline neutrino oscillations~\cite{DUNE}, could benefit from adapting the development presented in this work. Furthermore, our studies with detector variations offer insights that can inform the design and validation of future approaches in similar contexts.

\begin{ack}
This manuscript has been authored by FermiForward Discovery Group, LLC under Contract No. 89243024CSC000002 with the U.S. Department of Energy, Office of Science, Office of High Energy Physics. This report has the identification FERMILAB-CONF-25-0774-PPD. This work was made possible through the support of the Enrico Fermi Fellowships led by the Center for Spacetime and the Quantum, and supported by Grant ID \#63132 from the John Templeton Foundation. The opinions expressed in this publication are those of the author(s) and do not necessarily relect the views of the John Templeton Foundation or those of the Center for Spacetime and the Quantum.
\end{ack}


\bibliography{cite}

\begin{thebibliography}{27}
\providecommand{\natexlab}[1]{#1}
\providecommand{\url}[1]{\texttt{#1}}
\expandafter\ifx\csname urlstyle\endcsname\relax
  \providecommand{\doi}[1]{doi: #1}\else
  \providecommand{\doi}{doi: \begingroup \urlstyle{rm}\Url}\fi

\bibitem[Machado et~al.(2019)Machado, Palamara, and Schmitz]{SBNProgram}
Pedro~AN Machado, Ornella Palamara, and David~W Schmitz.
\newblock {The Short-Baseline Neutrino Program at Fermilab}.
\newblock \emph{Ann. Rev. Nucl. Part. Sci.}, 69:\penalty0 363--387, 2019.
\newblock \doi{10.1146/annurev-nucl-101917-020949}.

\bibitem[Abi et~al.(2020)]{DUNE}
Babak Abi et~al.
\newblock {Deep Underground Neutrino Experiment (DUNE), Far Detector Technical Design Report, Volume II: DUNE Physics}.
\newblock 2 2020.

\bibitem[Drielsma et~al.(2021{\natexlab{a}})Drielsma, Terao, Domin{\'e}, and Koh]{SPINEML}
Francois Drielsma, Kazuhiro Terao, Laura Domin{\'e}, and Dae~Heun Koh.
\newblock {Scalable, End-to-End, Deep-Learning-Based Data Reconstruction Chain for Particle Imaging Detectors}.
\newblock In \emph{{34th Conference on Neural Information Processing Systems}}, 2 2021{\natexlab{a}}.

\bibitem[Adams et~al.(2019)]{MicroBooNEML}
C.~Adams et~al.
\newblock {Deep neural network for pixel-level electromagnetic particle identification in the MicroBooNE liquid argon time projection chamber}.
\newblock \emph{Phys. Rev. D}, 99\penalty0 (9):\penalty0 092001, 2019.
\newblock \doi{10.1103/PhysRevD.99.092001}.

\bibitem[Radovic et~al.(2018)Radovic, Williams, Rousseau, Kagan, Bonacorsi, Himmel, Aurisano, Terao, and Wongjirad]{MLReview}
Alexander Radovic, Mike Williams, David Rousseau, Michael Kagan, Daniele Bonacorsi, Alexander Himmel, Adam Aurisano, Kazuhiro Terao, and Taritree Wongjirad.
\newblock Machine learning at the energy and intensity frontiers of particle physics.
\newblock \emph{Nature}, 560\penalty0 (7716):\penalty0 41--48, 2018.
\newblock \doi{10.1038/s41586-018-0361-2}.
\newblock URL \url{https://doi.org/10.1038/s41586-018-0361-2}.

\bibitem[Abratenko et~al.(2025)]{ICARUSCalibration}
P.~Abratenko et~al.
\newblock {Calibration and simulation of ionization signal and electronics noise in the ICARUS liquid argon time projection chamber}.
\newblock \emph{JINST}, 20\penalty0 (01):\penalty0 P01032, 2025.
\newblock \doi{10.1088/1748-0221/20/01/P01032}.

\bibitem[Adams et~al.(2018)]{WireCellSP}
C.~Adams et~al.
\newblock {Ionization electron signal processing in single phase LArTPCs. Part I. Algorithm Description and quantitative evaluation with MicroBooNE simulation}.
\newblock \emph{JINST}, 13\penalty0 (07):\penalty0 P07006, 2018.
\newblock \doi{10.1088/1748-0221/13/07/P07006}.

\bibitem[Acciarri et~al.(2015)]{SBNProposal}
R.~Acciarri et~al.
\newblock {A Proposal for a Three Detector Short-Baseline Neutrino Oscillation Program in the Fermilab Booster Neutrino Beam}.
\newblock 3 2015.

\bibitem[Abratenko et~al.(2023)]{ICARUSInagural}
P.~Abratenko et~al.
\newblock {ICARUS at the Fermilab Short-Baseline Neutrino program: initial operation}.
\newblock \emph{Eur. Phys. J. C}, 83\penalty0 (6):\penalty0 467, 2023.
\newblock \doi{10.1140/epjc/s10052-023-11610-y}.

\bibitem[Yu et~al.(2021)]{DNNROI}
Haiwang Yu et~al.
\newblock {Augmented signal processing in Liquid Argon Time Projection Chambers with a deep neural network}.
\newblock \emph{JINST}, 16\penalty0 (01):\penalty0 P01036, 2021.
\newblock \doi{10.1088/1748-0221/16/01/P01036}.

\bibitem[Andreopoulos et~al.(2010)]{GenieUserManual}
C.~Andreopoulos et~al.
\newblock {The GENIE Neutrino Monte Carlo Generator}.
\newblock \emph{Nucl. Instrum. Meth.}, A614:\penalty0 87--104, 2010.
\newblock \doi{10.1016/j.nima.2009.12.009}.

\bibitem[Heck et~al.(1998)Heck, Knapp, Capdevielle, Schatz, and Thouw]{Corsika}
D.~Heck, J.~Knapp, J.~N. Capdevielle, G.~Schatz, and T.~Thouw.
\newblock {CORSIKA: A Monte Carlo code to simulate extensive air showers}, 2 1998.

\bibitem[Ronneberger et~al.(2015)Ronneberger, Fischer, and Brox]{unet}
Olaf Ronneberger, Philipp Fischer, and Thomas Brox.
\newblock U-net: Convolutional networks for biomedical image segmentation.
\newblock \emph{CoRR}, abs/1505.04597, 2015.
\newblock URL \url{http://arxiv.org/abs/1505.04597}.

\bibitem[He et~al.(2015{\natexlab{a}})He, Zhang, Ren, and Sun]{uresnet1}
Kaiming He, Xiangyu Zhang, Shaoqing Ren, and Jian Sun.
\newblock Deep residual learning for image recognition.
\newblock \emph{CoRR}, abs/1512.03385, 2015{\natexlab{a}}.
\newblock URL \url{http://arxiv.org/abs/1512.03385}.

\bibitem[Graham et~al.(2017)Graham, Engelcke, and van~der Maaten]{uresnet2}
Benjamin Graham, Martin Engelcke, and Laurens van~der Maaten.
\newblock 3d semantic segmentation with submanifold sparse convolutional networks.
\newblock \emph{CoRR}, abs/1711.10275, 2017.
\newblock URL \url{http://arxiv.org/abs/1711.10275}.

\bibitem[Drielsma et~al.(2021{\natexlab{b}})Drielsma, Terao, Dominé, and Koh]{uresnet3}
Francois Drielsma, Kazuhiro Terao, Laura Dominé, and Dae~Heun Koh.
\newblock Scalable, end-to-end, deep-learning-based data reconstruction chain for particle imaging detectors, 2021{\natexlab{b}}.
\newblock URL \url{https://arxiv.org/abs/2102.01033}.

\bibitem[Zhou et~al.(2018)Zhou, Siddiquee, Tajbakhsh, and Liang]{nestedunet}
Zongwei Zhou, Md~Mahfuzur~Rahman Siddiquee, Nima Tajbakhsh, and Jianming Liang.
\newblock Unet++: {A} nested u-net architecture for medical image segmentation.
\newblock \emph{CoRR}, abs/1807.10165, 2018.
\newblock URL \url{http://arxiv.org/abs/1807.10165}.

\bibitem[He et~al.(2015{\natexlab{b}})He, Zhang, Ren, and Sun]{resnet}
Kaiming He, Xiangyu Zhang, Shaoqing Ren, and Jian Sun.
\newblock Deep residual learning for image recognition, 2015{\natexlab{b}}.
\newblock URL \url{https://arxiv.org/abs/1512.03385}.

\bibitem[Ghosh et~al.(2021)Ghosh, Nachman, and Whiteson]{SciRobust3}
Aishik Ghosh, Benjamin Nachman, and Daniel Whiteson.
\newblock {Uncertainty-aware machine learning for high energy physics}.
\newblock \emph{Phys. Rev. D}, 104\penalty0 (5):\penalty0 056026, 2021.
\newblock \doi{10.1103/PhysRevD.104.056026}.

\bibitem[Kasieczka and Shih(2020)]{SciRobust2}
Gregor Kasieczka and David Shih.
\newblock {Robust Jet Classifiers through Distance Correlation}.
\newblock \emph{Phys. Rev. Lett.}, 125\penalty0 (12):\penalty0 122001, 2020.
\newblock \doi{10.1103/PhysRevLett.125.122001}.

\bibitem[{Baldi} et~al.(2025){Baldi}, {Campos}, {Weng}, {Geniesse}, {Tran}, {Kastner}, and {Biondi}]{SciRobustNhan}
Tommaso {Baldi}, Javier {Campos}, Olivia {Weng}, Caleb {Geniesse}, Nhan {Tran}, Ryan {Kastner}, and Alessandro {Biondi}.
\newblock {Loss Landscape Analysis for Reliable Quantized ML Models for Scientific Sensing}.
\newblock \emph{arXiv e-prints}, art. arXiv:2502.08355, February 2025.
\newblock \doi{10.48550/arXiv.2502.08355}.

\bibitem[Aurisano et~al.(2016)Aurisano, Radovic, Rocco, Himmel, Messier, Niner, Pawloski, Psihas, Sousa, and Vahle]{NOvAML}
A.~Aurisano, A.~Radovic, D.~Rocco, A.~Himmel, M.~D. Messier, E.~Niner, G.~Pawloski, F.~Psihas, A.~Sousa, and P.~Vahle.
\newblock {A Convolutional Neural Network Neutrino Event Classifier}.
\newblock \emph{JINST}, 11\penalty0 (09):\penalty0 P09001, 2016.
\newblock \doi{10.1088/1748-0221/11/09/P09001}.

\bibitem[Sirunyan et~al.(2018)]{CMSML}
A.~M. Sirunyan et~al.
\newblock {Identification of heavy-flavour jets with the CMS detector in pp collisions at 13 TeV}.
\newblock \emph{JINST}, 13\penalty0 (05):\penalty0 P05011, 2018.
\newblock \doi{10.1088/1748-0221/13/05/P05011}.

\bibitem[Sadowski et~al.(2017)Sadowski, Radics, Ananya, Yamazaki, and Baldi]{AntimatterML}
Peter Sadowski, Balint Radics, Ananya, Yasunori Yamazaki, and Pierre Baldi.
\newblock {Efficient antihydrogen detection in antimatter physics by deep learning}.
\newblock \emph{J. Phys. Comm.}, 1\penalty0 (2):\penalty0 025001, 2017.
\newblock \doi{10.1088/2399-6528/aa83fa}.

\bibitem[Shimmin et~al.(2017)Shimmin, Sadowski, Baldi, Weik, Whiteson, Goul, and S{\o}gaard]{JetML}
Chase Shimmin, Peter Sadowski, Pierre Baldi, Edison Weik, Daniel Whiteson, Edward Goul, and Andreas S{\o}gaard.
\newblock {Decorrelated Jet Substructure Tagging using Adversarial Neural Networks}.
\newblock \emph{Phys. Rev. D}, 96\penalty0 (7):\penalty0 074034, 2017.
\newblock \doi{10.1103/PhysRevD.96.074034}.

\bibitem[Baldi et~al.(2019)Baldi, Bian, Hertel, and Li]{NoVAML2}
Pierre Baldi, Jianming Bian, Lars Hertel, and Lingge Li.
\newblock {Improved Energy Reconstruction in NOvA with Regression Convolutional Neural Networks}.
\newblock \emph{Phys. Rev. D}, 99\penalty0 (1):\penalty0 012011, 2019.
\newblock \doi{10.1103/PhysRevD.99.012011}.

\bibitem[Aguilar-Arevalo et~al.(2009)]{BNB}
A.~A. Aguilar-Arevalo et~al.
\newblock {The Neutrino Flux Prediction at MiniBooNE}.
\newblock \emph{Phys. Rev. D}, 79:\penalty0 072002, 2009.
\newblock \doi{10.1103/PhysRevD.79.072002}.

\end{thebibliography}





\appendix

\section{Description of ICARUS Detector Variation Dataset}

We have made a version of the training dataset available for use as an example of an image identification problem in particle physics with a need for robustness against variations in the data performance relevant for scientific applications. It is publically accessible through Harvard Dataverse at this link: https://doi.org/10.7910/DVN/QNAEDV. In particular, we include the input frames (deconvolved waveforms, two-plane matching, and three-plane matching) detailed in section~\ref{sec:DNNROI} and true ionization depositions from an ICARUS simulation of neutrino interactions in the Booster Neutrino Beam (BNB)~\citep{BNB, GenieUserManual} with overlaid cosmic-rays~\citep{Corsika}. ICARUS has four TPCs, each with a front and middle induction plane. There are thus eight sets of frames in each simulated event. The frames are already down-sampled in the time dimension (by 4x) and have dimensions (2112 wires $\times$ 1024 ticks) on the front induction plane and (5600 wires $\times$ 1024 ticks) on the middle induction plane. In our implementation, we subdivided the frame along the wire direction into chunks (two on front induction and four on middle induction) to further reduce the image size. The floating point frames are saved as 16-bit numbers to save space (in our DNN for this paper, we use 32-bit floating points). We include a sample with the Nominal detector simulation, and the OmniDetector simulation detailed in table~\ref{tab:detvar}. We also include, as smaller validation samples, particular detector simulations with especially challenging detector conditions: high noise (120\%), low electron lifetime (\SI{2.5}{\milli\second}), and maximal middle induction intransparency (as observed in ICARUS).

\end{document}